\documentstyle{article}

\begin{document}
\title{Comment on "Effect of entanglement on the decay dynamics of a pair of H(2p) atoms due to spontaneous emission"}
\date{}
\author{Pedro Sancho (a), Luis Plaja (b) \\ (a) Centro de L\'aseres Pulsados, CLPU.  E-37008, Salamanca, Spain \\ (b) Area de Optica. Departamento de F\'{\i}sica Aplicada \\ Universidad de Salamanca. Pl. de la Merced s/n, 37008, Salamanca, Spain}
\maketitle
\begin{abstract}
Tanabe et al (Phys. Rev. A {\bf 82} 040101(R) 2010) have
experimentally demonstrated that the emission properties of unstable
atoms in entangled and product states are different. The authors
define an apparent decay time as a fitting parameter which falls
below the lifetime of the single atom for entangled pairs. We argue
that their results about coincidence time spectra are correct, but
those concerning  lifetimes cannot be considered conclusive because
they assume the emission of photons by the two atoms to be
independent processes, a doubtful hypothesis for entangled states. We
suggest an improved evaluation of the lifetimes based on a rigorous
approach, which demands some modifications of the experimental
procedure.
\end{abstract}
\vspace{4mm}

PACS: 03.65.Ud; 32.70.Cs; 33.80.-b

\vspace{4mm}

The presence of entanglement can modify the emission properties of a
pair of unstable atoms. In \cite{tan} it has been experimentally
demonstrated that the coincidence time spectra of two photons
emitted by $H$ atoms in the unstable $2p$ state, generated in the
photodissociation of a $H_2$ molecule, are different when the atoms
are entangled. In particular, when the experimental data are fitted
to an exponential distribution the apparent decay time coefficient in
the case of entangled states is approximately half the value of
product states. This is a fundamental result, leaving no room for
doubts on the differences of the emission properties in both cases.

Usually the decay time is identified with the lifetime of the
system. In this line, the authors of \cite{tan} present a simple
reasoning trying to define an apparent decay time of the single
atoms in the entangled pair. They compare their experimental
distribution with the one expected for a pair of particles emitting
independently (in a product state), concluding that the lifetime of
single atoms is apparently half of a single atom. However, it seems
doubtful that the condition of independent emissions can be
translated to the case of entangled states (in whose case the
apparent decay time defined in \cite{tan} would be rigorously the
actual decay time). Then the results in \cite{tan} cannot be
considered as a proper evaluation of the decay times. One should use
an evaluation of the lifetimes free of that problematic assumption.

A rigorous method to calculate the lifetimes of single atoms in
entangled states was presented in \cite{sp}, where the possibility
of a different behavior for the emission properties of entangled and
product states was also suggested. To our opinion, the method can be
easily adapted to the experiment in \cite{tan}. The temporal
variation of the number of entangled pairs, $n_e$, is given by
$dn_e/dt=-\Gamma _f n_e$, where $\Gamma _f$ is the emission rate of
the first photon by the entangled two-atom system. After this first
emission one of the atoms is in the ground state and the
two-particle state becomes a product one. Next, we denote by $n_i$
(the index $i=A,B$ labels the two atoms) the number of unstable
atoms of type $i$ in a product state. Their variation rates are
given by $dn_i/dt=(\Gamma _f/2) n_e -  \Gamma _s n_i$. The first
term, the source term, represents the generation of unstable atoms
in product states from the initial entangled state. The coefficient
$1/2$ comes from the fact that being atoms $A$ and $B$ identical,
the probabilities of the first decay process leaving an unstable
atom of type $A$ or $B$ are equal. On the other hand, the second
term corresponds to the decay of the unstable atom to the ground
state via the second photon emission. In the usual approximation of
time independent emission rates, one can easily obtain $n_e(t)= n_0
e^{-\Gamma _f t}$, with $n_0$ the initial number of entangled pairs
(we take the initial time $t=0$ as that of the formation of the
entangled pair), and $n_i(t)= n_0\Gamma _f (e^{-\Gamma _f
t}-e^{-\Gamma _s t})/2(\Gamma _s - \Gamma _f)$. The equation for
$n_i$ shows that in the case of entangled states one cannot express
the relevant variables as the product of two exponentials, as it is
assumed in Eq. (4) in \cite{tan}, to define the apparent decay time.
The impossibility of expressing the variables that way reflects the
inadequacy of trying to translate the condition of independent
emissions from the product to the entangled case.

The total number of unstable atoms of type $i$ is $n_e(t)+n_i(t)$,
that is, the sum of unstable atoms in entangled and product states.
By definition, the lifetime $\tau $ ($\tau _A = \tau _B$) is given
by the relation $n_0 e^{-1}=n_e (\tau )+n_i(\tau )$. Note that the
initial number of unstable atoms of type $i$ coincides with the
initial number of entangled pairs. Combining all the previous
expressions we have
\begin{equation}
\frac{2(\Gamma _s - \Gamma _f)}{e}=(2\Gamma _s - \Gamma _f) \exp (-\Gamma _f \tau) - \Gamma _f \exp (-\Gamma _s \tau )
\label{eq:uno}
\end{equation}
Equation (\ref{eq:uno}) shows that the lifetime of the single atoms
can be calculated once $\Gamma _f$ and $\Gamma _s$ are
experimentally determined. At variance with \cite{tan}, the lifetime
of single atoms does not depend only on one parameter (their {\it
decay time} coefficient) but on two ($\Gamma _f$ and $\Gamma _s$).

From the experimental point of view, one does not measure $n_e(t)$
and $n_i(t)$ but the temporal distributions of emitted photons. The
distribution of first emitted photons, $N_f$, can be easily
determined from the obvious condition $n_e(t)+N_f(t)=n_0$, giving
$N_f(t)=n_0(1-e^{-\Gamma _f t})$. This distribution gives the number
of first-type photons emitted between the formation of the entangled
pair, $t=0$, and the time $t$. The distribution of photons emitted in the
second place, $N_s$, can be deduced from the relations
$n_e(t)+n_i(t)+n_i^g(t)=n_0$ and $2n_i^g (t)=N_f(t)+N_s(t)$, where
$n_i^g$ is the number of atoms of type $i$ in the ground state. The
coefficient $2$ in the last expression is due again to the fact that
both atoms are identical. We have the distribution $N_s(t) =N_f(t)-
n_0 \frac{\Gamma _f}{\Gamma _s -\Gamma _f}(e^{-\Gamma _f
t}-e^{-\Gamma _s t})$. From the expressions for $N_f$ and $N_s$ one
can deduce $\Gamma _f$ and $\Gamma _s$ by fitting the experimental
data.

Next, we derive the coincidence time spectra in our approach,
showing that it agrees with the one obtained in \cite{tan}. The
coincidence time spectra measures the number of double-emission
processes taking place for each time separation. The only relevant
temporal variable is the time separation $|\Delta t|=t_s - t_f$,
becoming irrelevant $t_f$ and $t_s$ as independent variables.
Physically, we only care about the separation of the emission
events, regardless of the time elapsed after the generation of the
entangled pair. Then the quantities $n_i$ and $N_i$ must be
expressed in terms of $|\Delta t|$ instead of $t_f$ and $t_s$. We
denote by $\tilde{n}$ the total number of unstable non-entangled
atoms ($\tilde{n}=n_A + n_B$). Its temporal variation is obtained
adding those of $n_A$ and $n_B$ and taking into account that the
source term is null (by definition, at $|\Delta t|=0$ all the
entangled pairs decay to product ones): $d\tilde{n}/d|\Delta
t|=-\Gamma _s \tilde{n}$. Note that we assume $\Gamma _s$ to be
time-independent. The solution of this equation is
$\tilde{n}(|\Delta t|)=n_0 \exp (-\Gamma _s |\Delta t|)$ (the total
number of unstable non-entangled atoms at $|\Delta t|=0$ is $n_0$).
The number of coincidence counts in a small interval $\delta |\Delta
t|$ around $|\Delta t|$ is given by the change $\delta
\tilde{n}(|\Delta t|)=\tilde{n}(|\Delta t|+\delta |\Delta
t|)-\tilde{n}(|\Delta t|)$ of unstable non-entangled atoms in that
interval. Using the usual Taylor's expansion we have $|\delta
\tilde{n}(|\Delta t|)| \approx (\Gamma _s n_0/2)\exp (-\Gamma _s
|\Delta t|)\delta |\Delta t|$. Thus, we have that the number of
coincidence counts for the time separation $|\Delta t|$ is
proportional to $\exp (-\Gamma _s |\Delta t|)$. This distribution is
equivalent to that obtained in \cite{tan} with $\Gamma _s =1/\tau _{app} $,
where $\tau _{app} $ is the apparent lifetime.

The previous reasoning shows that the results in \cite{tan}
concerning the coincidence time spectra are not modified in the more
general framework considered in this Comment. Our statement can be
tested determining the coincidence time spectra in an experiment
where the measurement of $t_0$ is done. On the other hand, it would be
interesting to have some hints on the expected values of the
lifetimes in our approach. We note that the emission rate of the
first photon coincides with the disentanglememnt rate $\Gamma
_{dis}$, which can be evaluated provided the evolution operator of
the complete system is known. An example of such an evaluation was
presented in \cite{sp}. For the system considered in \cite{tan} the
evaluation would be much more difficult because, at variance with
\cite{sp}, the alternatives for the emission of the photon causing
the disentanglement are undistinguishable, giving raise to
interference effects. However, we can provide a lower bound for
$\Gamma _{dis}$. In \cite{yu} it was shown that entanglement decays
at least as fast as the sum of the individual decoherence rates. As
in the Born-Markov approximation these decoherence rates are half
the spontaneous decaying rates we have $\Gamma _f =\Gamma _{dis}
\geq (\Gamma _A + \Gamma _B)/2=\Gamma _0$, where $\Gamma _0$ denotes
the spontaneous decaying rate of a free (in a product state) atom.
Moreover, we have from \cite{tan} that $\Gamma _s \approx
2\Gamma _0$. Combining both results with Eq. (\ref{eq:uno}), we have
$\tau =1.31 \tau _0$ (for $\Gamma _f =\Gamma _0$ and $\Gamma
_s=2\Gamma _0$), with $\tau _0=1/\Gamma _0$ the lifetime of the free
atom. This provides an upper limit for the lifetime of atoms
initially in entangled states, $\tau \leq 1.31 \tau _0$. Other
representative values are, for instance, $\tau =0.79 \tau _0$ for
$\Gamma _{dis}=2\Gamma _0$ (using L'H\^opital's rule) or $\tau =0.33
\tau _0$ for $\Gamma _{dis}=8\Gamma _0$.

We note however, that our suggestion requires a precise knowledge of
the time of formation of the entangled pair, which is used above to
set the temporal origin ($t=0$).  In \cite{tan} this step was not
necessary because the definition of the apparent lifetime needs only
the difference of times, $t_s  -  t_f$. This is in principle
feasible with the photon source considered in Tanabe's paper, with
pulse lengths about ten times smaller than the lifetime of H(2$p$),
and of the same order of the apparent lifetime of the entangled
pair. However, a more precise determination may require the use of
shorter UV pulses generated either by harmonic selection of a
multicycle pulse  \cite{sekik04A} or by Fourier synthesis of the
higher energy part of the high-order harmonic spectra generated by
intense few cycle pulses \cite{ferra10A}.

\end{document}